\documentclass[12pt]{article}
\usepackage[dvips]{graphicx}
\usepackage[hypertex,linkcolor=red]{hyperref}

\newcommand{\be}{\begin{equation}}
\newcommand{\ee}{\end{equation}}
\newcommand{\bel}[1]{\begin{equation}\label{#1}}
\newcommand{\bea}{\begin{eqnarray}}
\newcommand{\eea}{\end{eqnarray}}
\newcommand{\beal}[1]{\begin{eqnarray}\label{#1}}
\newcommand{\nn}{\nonumber}
\newcommand{\nin}{\noindent}
\newcommand{\nfn}[1]{\renewcommand{\theequation}
           {#1.{\arabic{equation}}}\setcounter{equation}{0}}
\newcommand{\as}{\alpha_s}

\def\d{\partial}

\def\Pm{\hbox to 5pt {$I\!\!\!\!P$}}
\def\Pn{\hbox to 15pt {$I\!\!\!\!P_n$}}
\def\P{I\!\!\!\!\!P}
\def\P1{\hbox to 10pt {\hfill \large $\cal P$} }
\def\F{\hbox to 10pt {\hfill \large $\cal F$} }
\def\V{\hbox to 10pt {\hfill \large $\cal V$} }
\def\FC{\hbox to 20pt {\large $\cal FC$ } }
\def\Fd{\hbox to 18pt { {\large $\cal F$} \hss {\it~d} } }
\def\dlr{\hbox{\raisebox{8pt}{$\leftrightarrow$}$\!\!\!\!\!\d$}}

\begin{document}

\title{\bf On the Multi-Component Pomeron in high energy
   hadronic interactions}

\author{ {\sc O.V. Kancheli}\thanks{E-mail: kancheli@itep.ru} \\
     {\it  Institute of Theoretical and Experimental Physics, }  \\
     {\it  B. Cheremushinskaya 25, 117 259 Moscow, Russia. } }
\date{}
\maketitle

\begin{abstract}
We consider  the phenomenon of regge cut splitting, corresponding
to the BFKL pomeron into an infinite sequence of regge-poles \Pn,
which happens when one takes into account the running of QCD
coupling $\alpha_s (u)$ with scale. The pomeron levels
~$\Pm_n,~~~n=1,2,...\infty$ have intercepts $j_n(0) \simeq 1+c/n$
and represent the BFKL-like objects with different mean
virtualities $u_i \simeq \ln p^2_{\bot}/\Lambda^2 \sim n$ on
internal lines of corresponding gluon ladders. The first members
of the $\Pm_n$ sequence, after adjusting their parameters, may
effectively include the nonperturbative part of pomeron, and this
way all \Pm ~components - soft and hard may be treated in a unique
way. We also  illustrate how the \Pn can enter into the
phenomenological regge description of some high energy processes.

\end{abstract}
\newpage
\setcounter{footnote}{0}
\section*{\bf  Introduction.}
\nfn{1}

Regge phenomenology based on the reggeon diagrams approach
\cite{grd} allowed to explain and to describe quantitatively
various high energy hadronic phenomena and pomeron is one of the
main ingredients of such an approach.

 In QCD pomeron is usually considered as a specific gluon ladder
equipped with some nonperturbative interactions. The BFKL
construction \cite{BFKL},\cite{fl} refined this in a double
logarithmic approximation - here pomeron is a ladder (bound state)
of two reggesed gluons. In this approximation QCD coupling
$\alpha_s$ is not a running one, and this leads to the transverse
conformal-invariance of the parton-gluon motion. As a result,  a
cut in j-plane at $j = 1 + c \alpha_s$, and not a pole corresponds
to pomeron. The mean virtuality of gluons  such a pomeron is
composed of are not small, and these partons (due to the
conformal-invariance) diffuse freely in virtuality $u \simeq \ln
p^2_{\bot}/\Lambda^2$ with growth of rapidity.

 But when one takes into account the running of $\alpha_s(u)$ with
scale and supplements it with rather general boundary conditions
in the infrared region, so that these conditions represent
confinement (in fact one stops to move partons to lower $u$), then
the pomeron cut transforms into a sequence of regge poles \Pn with
intercepts $j_n(0) \simeq 1+c/n$ ~\cite{Lip86}-\cite{ekr}.

It is essential that the mean virtualities of gluons, from which
the~\Pn  ladders are constructed, grow with $n$ like $<u_n> \equiv
<\log k_{n\bot}^2/\Lambda^2> \sim n$.
 The first member of this sequence $\Pm_1$ is almost soft, and
its properties essentially depend on the details of infrared
boundary conditions. Therefore it is possible to expect, that by
adjusting these conditions, or directly varying the parameters of
the first poles one may effectively take into account the
nonperturbative properties of the soft pomeron.
 There exists a number of reasons that can support such a hope.

It is well known that one can quite good approximate almost all
``needed'' properties of pomeron (so to explain the most of
existing soft high energy data)  by simple ladder diagrams with
known hadrons $(\pi, \rho,...)$ , where high transverse momenta
are cut by vertices (by some form-factors consistent with the low
energy hadronic physics). All such models lead to pomeron as a
regge pole with some nonlinear trajectory.  ~One would expect the
same type answer, while trying to reconstruct the pomeron
structure from dominant multipherical processes using unitarity
and analyticity in all channels.
  This suggests that one may take into account all essential
nonpertutbative mechanisms using the multiperipheral ladder
dynamics with the adequate choice of pole trajectories and regge
vertices.

One may take a slightly different approach and consider the Fock
type wave function of fast hadron in the light-cone variables
using QCD quark-gluon partons - this is what is contained in BFKL
and their generalizations. In this case all nonperturbative
effects are connected or with the zero-mode partons in the wave
function or with interaction in the final state (after collision)
when various QCD strings may be created and decay. These
zero-momentum partons interact with other partons (with the
arbitrary rapidity) and can essentially modify the structure of
the parton wave function at low transverse momenta. In principle
one may integrate out these zero-modes, and this induces various
additional soft multigluon terms in effective QCD light-cone
Hamiltonian. Some of these terms may also correspond to the
effective QCD strings stretched between ladder gluons if their
transverse separation becomes larger than some critical value.
These effects are approximately local in rapidity, due to local
color neutralization in the gluon ladder within some interval of
rapidity depending on their virtuality. Therefore it seems that
the nonperturbative effects shall not change the pole structure of
\Pm ~after such a dressing of the ladder structure.

The other type of reasoning comes from the consideration of high
energy scattering in dual models, where the pomeron appears
naturally as a cylinder type construction and where the string
degrees of freedom already represent the main QCD nonperturbative
effects. When such a dual model is considered in $AdS_5$ space
\cite{posu},\cite{bpst} one may at once consider also all main
perturbative contributions.
 If the $AdS_5$ profile is deformed so to reproduce the $\alpha_s$
running then the attraction of strings to infrared region shall
take place and the conformal \Pm ~splits into \Pn - like states.
Here the higher \Pn states appear as the same sting type state as
the soft one, but placed on average at a larger 5-th coordinates,
where they become smaller and more virtual.

So one can expect that such a multiple \Pm ~structure should
represent the main perturbative and nonperturbative QCD effects
and can be used for universal phenomenological description of
various processes in rather large ranges of energy and virtuality
\footnote{ On can imagine a slightly different possibility that
because of the nonlinear effects in  parton evolution and at not
too large rapidity the continues regge cut (like in the simple
BFKL case) will be not a worse approximation, and here we have the
continuous spectrum of pomerons  \textbf{P}(u) , labelled by their
mean virtuality. Approximately  $\Pn \simeq  \textbf{P}(u)$ for $n
\simeq c u$, but the multireggeon vertices corresponding to these
states may differ significantly. Such type approach to soft and
hard processes with one continuous pomeron was introduced recently
in~ \cite{rmk}. }.~
 The experimental data support the multicomponent structure of \Pm
~and a number of models with such a \Pm ~where proposed. The
simplest model contains pomeron with two components - soft and
hard one \cite{ld}. Models with more component were also
introduced, and they lead (see for example \cite{pp}) to good fits
for the existing data.

 \vspace{2mm}
In this paper we discuss some aspects of these problems. We remind
the arguments leading to a multi ~\Pn pole structure of pomeron
and try to elucidate the physics of this phenomenon.

\subsection*{~1.  The ~\Pn ~states from BFKL equation
            with running~$\alpha_s$}
\nfn{1}

The infinite sequences of regge poles condensed to the position of
born contribution naturally arise in theories where the strength
of interaction weakly depends on scale. They correspond to the
effective perturbative ``ladders'' with different mean internal
virtuality. ~Such a phenomenon arises in the asymptotically free
$\lambda \varphi^3_6$ theory \cite{Lovelace} and in QCD
~\cite{Lip86}~. ~The intercept of these poles at large $n$ is
$j_n(0) \simeq j_{Born}(0) + c/n$, where $j_{Born}(0)=-2$ for
$\lambda \varphi^3_6$ and $j_{Born}(0)=1$ for QCD. ~The
coefficient $c$ defines the density of these poles ; $c^{-1} \sim
b$ - the first coefficient of the $\beta$ function of theory which
defines how fast the parton coupling runes with scale. For the
first $\Pn$ states (these components of pomeron, give the main
contribution to soft processes) their perturbative intercepts
$j_n(0)$ may be essentially shifted due to a nonperturbative
effects.

\vspace{5pt}

Here we briefly remind how such pomeron states \Pn appear in the
BFKL parton evolution. For the nonrunning $\alpha_s$ there is a
transverse scale invariance - as a result we receive the
continuous spectrum of regge pole states - the BFKL cut from $j(0)
-1 ~\sim \alpha_s$~,~ representing the upper border of spectrum.
For running  $\alpha_s$ this cut ``splits'' into an infinite
system of poles accumulating at the point $j(0) =1$.
 To see it one may start from the simplest generalization of the
BFKL equation to the case of the running coupling  $\as$~:
 \beal{BFKLrunning}
\frac{1}{\as(u)}\frac{\partial
  f(y,u)}{\partial y}) ~=~   \int d u_1  L(u_1-u)
  f(y,u_1)~~,~~~~~~~~~~~~
     \\
L(u) ~=~  \int^{i \infty}_{-i \infty} ~
   {\cal L}(\beta) ~e^{u \beta}
 ~\frac{d\beta}{2\pi i} ~~,~~~~~~~~~~~~~~~~~~~~~ \nn
     \\
~~~{\cal L}(\beta) = \psi(\beta) + \psi(1-\beta) - 2 \psi(1)~~,~~
~~~~u \equiv \log k^2_{\bot}/\Lambda^2
   \nn
\eea
Then one can approximate the kernel of such an equation by its
regular part. This corresponds to the expansion of the kernel
${\cal L}(\beta)$ in $\beta =\d/\d \log k^2_{\bot}$ - Laplace
conjugated to $u$ around the minimum  ~~${\cal L}(\beta) \simeq
{\cal L}(1/2) + (1/2) (\beta - 0.5)^2 {\cal L}''(0.5) +\ldots $,
and taking only first terms. Then we have from
(\ref{BFKLrunning}):
 \bel{bfkl_dif}
\frac{1}{\as(u)}\cdot
\frac{\partial f}{\partial y}  ~=~ \delta~f + B
\frac{\partial^2 f}{\partial u^2}  ~,
 \ee
where for standard QCD with running $\as(u) \simeq (a+bu)^{-1}~$ ,
$a \simeq~ 0.6$~, ~$b~\simeq~0.7$~, $\delta~=~4N_c~\ln 2/\pi
\simeq 2.6$~,~~~ $B = 14 N_c ~\zeta(3)/\pi  \simeq 16$~.

The equation (\ref{bfkl_dif}) for the gluon density $f(y,u)$ has
the form of a diffusion equation with the parton branching, in
which the branching coefficient $\as (u) \delta$ and the diffusion
coefficient $\as (u) B$ depend on the ``coordinate" $u$. In the
same language rapidity plays the role of the time.
 This analogy is useful for understanding of the
behavior of the function $f(y,u)~$, and it will be used later.

Note that the nonsingular part of kernel at $\beta \simeq 1/2$
corresponds to a parton chain evolution where the changes of u at
individual steps are not large :  ${\bf \delta} u \sim \as$.~
 The large jumps of $u$ correspond to the region $\beta
\sim 1/u$ where ${\cal L}(\beta) \sim 1/\beta$~, and for which
${\bf \delta} u \sim \as u$~.

Going to the complex angular momenta in (\ref{bfkl_dif}) as
$f_{\omega} = \int e^{y \omega} f$~,~ $j=1+\omega$ we come to the
equation
 \bel{de_o}
  (a +b u) \cdot \omega f_{\omega}
   = \delta f_{\omega} +
     B \frac{\partial^2 f_{\omega}}{\partial u^2} ~,
 \ee
which has the Airy form. Its solution  can be written as
 \bel{eiri1}
  f_{\omega}(u)  = f_{\omega}(u_0) \cdot
       \frac{Ai ( ~z(u,\omega)~ )}{Ai( ~z(u_0,\omega)~ )}~,~~~
       z(u,\omega) =  ~\Big( u +
          \frac{a}{b} - \frac{\delta}{b\omega}\Big)
    \Big(\omega\frac{b}{B}\Big)^{\frac{1}{3}}   ~,~~
 \ee
where $Ai(z)$ is the Airy function, and where  boundary conditions
at $u = u_0$~  are defined by the function $f_{\omega}(u_0)$ close
to the infrared region
\footnote{Here we reproduce the reasoning given in \cite{jlo}}.

In the solution (\ref{eiri1}) the gluon dynamics is, in fact,
divided between two factors. One of them is the function
$f_{\omega}(u_0)$ ~-~ its singularities  in $\omega$ represent the
``soft'' part of the pomeron coming from the region $u <u_0$ and
are  mainly generated by the nonperturbative mechanisms. The
factor $Ai (z(u,\omega))/Ai(z(u_0,\omega))$~, oppositely,
represents the hard part of pomeron. The zeros of denominator
$Ai(z(u_0,\omega))$ in $\omega$ are approximately located at
 \footnote{The entire function $Ai (z)$ has zeros only at negative
$z$ at points $z \simeq -2.33, ~-4. , ~-5.5 ,...$. Its asymptotic
form at $-z \gg 1$ is  $Ai (-z) \simeq  \pi^{-1/2} z^{-1/4} \sin
(\frac{2}{3} z^{3/2} + \pi/4 )$ . }
 points
 \bel{ppos}
 \omega_n ~=~ \frac{C_1}{n+C_2}~+~O(1/n^2),~~~~~ n=1,2,...
 \ee
$$
      C_1 = \frac{2}{3\pi b} \Big(\frac{\delta}{B}\Big)^{1/2}~\simeq
    0.45  ~~,~~~C_2 \sim 1
$$
The poles with $n \sim 1$ correspond to relatively soft
processes~; they become harder and harder when  n grows. The order
of mean logarithms of a transverse momenta (virtuality) in BFKL
ladder, corresponding to $n$-th pole, are
$$
 <u_n>  ~\simeq~  c_3~ n~~,~~~~~~~
   c_3 \simeq \frac{3\pi}{4} \Big(\frac{B}{\delta}\Big)^{1/2}
   ~\simeq 5
$$
and they grow  fast with n. The distribution of $u$ in \Pn states
around the mean value $<u_n>$ is wide , also of order of $<u_n>$.
This may be directly seen from the solution (\ref{eiri1}).

It is essential that the $O( 1/n^2)$ and  higher corrections to
$\omega_n$ in (\ref{ppos}) are $u_0$-dependent (see~(\ref{omn})) -
these terms define the slope of ~\Pn trajectories. For the first
poles (minimal $n=1$ , and partially $=2$), these corrections,
also depending on the boundary condition at $u_0$, may in
considerably change the $\sim C_1/n$ positions of ~$\Pn$. This
enables to adjust the values $f_{\omega}(u_0)$ and $u_0$ in such a
way that the $\Pm_1$ coincides with the ``soft'' pomeron, and in
this way take into account all main nonperturbative effects.

\vspace{3mm}

\subsection*{~2.~Simple quantum-mechanical model for
the \Pn ~states}
 \nfn{2}

The following analogy is useful to make the structure of the ~\Pn
states more evident \cite{jlo}. If we change  $y \rightarrow iy$
in  Eq.(\ref{bfkl_dif}) we become the Schroedinger equation for
one-dimensional motion of a ``particle'' with the coordinate $u$
varying in the interval $u_o < u < \infty$ and $y$ playing the
role of time. The Hamiltonian of this motion, as follows from
Eq.(\ref{bfkl_dif}), is given by
 \bel{efham}
    E ~=~ V(u)  +~ {\hat p}^2/2 m(u)~~,
 \ee
where the momentum $~~{\hat p} = i {\hat \beta} =
-i\partial/\partial u~$.~ The potential and the u coordinate
dependent  mass in (\ref{efham}) are :
 \bel{pot_mass}
   V(u) = -\delta \as(u) \simeq  - \delta / (a+bu) ~;~~~~~
   m(u) = \left( 2 B \as(u) \right)^{-1} \simeq
   (a + bu)/ 2 B
 \ee
 In a case of a constant $\as(u)$  we have a free motion in $u$.
So, when we put the corresponding ``particle"(it is the Pomeron
ladder) at initial time $y=0$ to some position $u_1$ (it is with
initial \Pm ~~transverse size $\sim \exp (-u_1)$) then at the
later time there will be simply a spreading of a wave packet,
described by the Green function of free motion - it corresponds to
the ``standard'' BFKL behavior (Gaussian spreading in virtuality
u).

 But for the running $\as(u)$  the motion is not free - we have long
range (in $u$) attractive forces $\sim 1/u^2$, acting to the
direction of small $u$.~
\begin{figure}[h]
\begin{center}
\includegraphics[scale=0.8,keepaspectratio=true]{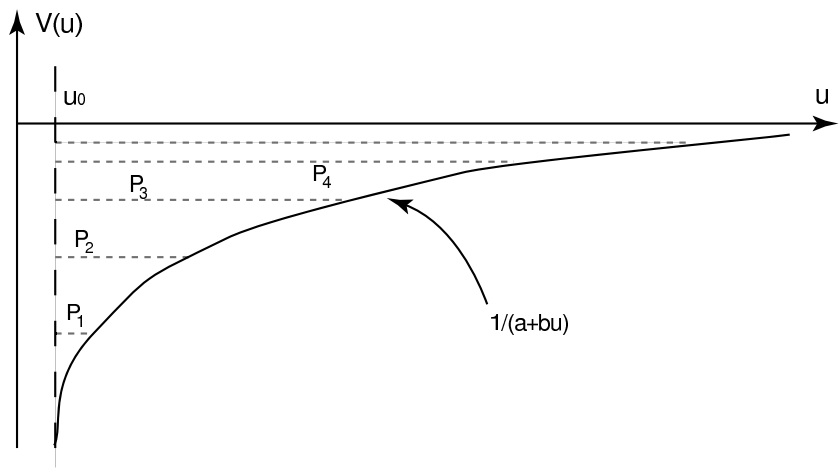}
\caption{\small Qualitative form of potential $V(u)$
 and of the ~$\Pn$ levels.
} \label{Fig1}
\end{center}
\end{figure}

In this potential, due to one-dimensionality of the motion, the
infinite ``Coulomb-like" series of bound states exist. For the
high lying levels their motion is on average located at large
``distances'' ~$ u \gg u_0$~ where the motion is quasiclassical.
Solving the expression (\ref{efham}) for $E$ relative to $p$ we
can write the standard quasiclassical quantization conditions
 \bel{qcon}
 n~\pi ~=~  I_n~,~~~
 I_n ~=~ \int_{n_0}^{u_{max}(\varepsilon)} ~p(u)~d u~,~~
 p^2 = 2 m(u)(\varepsilon-V(u))~,
 \ee
where $u_{max}(\varepsilon) ~=~ (\delta - a \varepsilon)/b
\varepsilon$.~~~ Than, using (\ref{pot_mass}) we find from here
$$
I_n ~=~ \pi~ \frac{\omega_0}{\varepsilon}~
  \Big( 1 - \varepsilon ~\frac{a+b ~u_0}{\delta}  \Big)^{3/2}~,~~~
  \omega_0 ~=~ \frac{2}{3\pi}  \Big( \frac{\delta}{b} \Big)
  \Big( \frac{\delta}{B}\Big)^{1/2}~,
$$
and the equation for the position of $\varepsilon$-levels takes
the form
 \bel{elev}
\varepsilon ~=~
 ~~=~\frac{\omega_0}{n}~
  \Big( 1 - \varepsilon ~\frac{1}{\delta \alpha_s(u_0)}  \Big)^{3/2}
  \ee
From here for large $n$ we find the simple solution for positions
of ~$\Pn$ levels
 \bel{omn}
 \varepsilon ~\simeq~ \omega_0
 ~\frac{\alpha_s(u_0)}{n \alpha_s(u_0) + C_2}
 ~~=~~\frac{\omega_0}{n~ +~ C_2(a+bu_0) }~~,~~~~
 C_2 = \frac{3\omega_0}{2\delta}~,~~~~
 \ee
which has the same structure as (\ref{ppos}). But even for minimal
$n=1$ this estimate may be acceptable, because parameters entering
(\ref{omn}) for realistic QCD are approximately
$$
 \omega_0 ~\simeq~ 0.3~~~,~~~~  C_2 =  \simeq 0.172~.~~~
$$
The value of $\alpha_s(u_0) ~\simeq~ 0.3 \div 0.6$~,~ entering
(\ref{omn}) should be taken at such $u_0 \sim 2\div 3$ ~,~ when we
go to strong nonperturbative dynamics and $\alpha_s(u_0)$ become
frozen.
 So, if we choose the value of the nonperturbative border at $\Lambda_0
\simeq 0.5 \div 1 ~GeV$, and $\Lambda_s \simeq 0.15 GeV $~, which
corresponds to $u_0 \simeq  3 \div 5$, we become from (\ref{omn})
for $\Pn$ intercepts the estimate ~$\omega_1 \simeq
0.2$~,~~$\omega_2 \simeq 0.12$,~~ $\omega_3 \simeq 0.085$~,~...

\nin The wave functions $\psi_n(u)$ of $P_n$ states
  \bel{pvf}
 \psi_n(u) ~\sim~  \frac{1}{\sqrt{n}} \exp
    \Big( ~i\int p(u)~du \Big)
   \ee
are spread at large n for large interval of virtualities $\sim u
\sim n$, inside the potential well. But $\psi_n(u)$ are rather
smooth and even at large n are influenced by the ``repulsive''
boundary conditions at $u_0$.

Therefore one may try to adjust the potential $V(u)$ near $u_0$ in
such a way to ``move'' the nonperturbative contributions in the
pomeron state from $\phi_{\omega}$ directly into ~$\Pm_1$. After
that all singularities of $\phi_{\omega}$ in (\ref{eiri1}) are
located at $\omega \le 0$~.

Because confinement (as we are understanding it now) must strongly
restrict the color particles motion at large distances we simply
suppose that effective potential $V(u)$ becomes very big and
positive at $u<u_0$. It means that the levels \Pn will always
remain discreet, and only the position of few first terms \Pn must
by adjusted to take into account nonperturbative effects.

From (\ref{elev}) it is simple to find the slope of \Pn
trajectories defining them from the reaction of levels on the
shift of $u_0$ boundary conditions

$$
\alpha'_n (0) ~\simeq~
    -\frac{1}{\Lambda_0^2} ~\frac{\d \varepsilon_n}{\d u_0}
    ~\simeq~ \frac{1}{\Lambda_0^2} \frac{C}{(n)^2}
    ~~~,~~~~ C \simeq  0.25 ~~,~~~
    u_0 \equiv \log \Lambda_0^2/\Lambda_s^2
$$
With the same method we can find the full regge trajectories for
$\Pn$, associating the shift of boundary from $u_0 = \log
(\Lambda_0^2/\Lambda_c^2$ to the new  boundary $\log
((\Lambda_0^2+|t|)/\Lambda_c^2$  with growth of $|t|$ from the
zero value. The result is~:
 \bel{pntr}
  j_n(t) ~\simeq~ 1 ~+~
  ~\frac{\omega_0 }{n + C_2~ (\alpha_s(\hat{u}))^{-1} }
  ~~,~~~~~\hat{u} = \log \frac{\Lambda_0^2+|t|}{\Lambda_c^2}
 \ee
Such an expression appears because the position of Pomeron pole at
some $-t$ is approximately the same as its position, when the
region of the transverse motion of intermediate gluons is
restricted by the condition $~u > \log{(-t)}$~. Note, that for
large $u_0$ the region of motion $u_{max}(n) -u_0$ is concentrated
at large $u$ also for $n=1$, and therefore  the corrections to
quasiclassical expressions for $\alpha_n(t)$ may be small for
large $-t$ even for the ground level $\Pm_1$.

\vspace{3mm}

\subsection*{~~3.~~ The higher order corrections to \Pn ~states}
\nfn{3}

The BFKL equation is formally valid in rather specific conditions,
corresponding to the main logarithmic approximation in $\alpha_s
\log s$. One may expect that the higher in $\alpha_s$ corrections
and the nonperturbative contributions can induce additional
properties of pomeron not seen in lowest approximation when only
running of $\alpha_s$ is taken into account. They may be~:
\vspace{1mm}

\nin * ~Corrections to \Pn parameters and the mixing of \Pn with
      multigluon states.

\nin *~  Nonperturbative effects.

\nin  *~ Nonlinear terms in the gluon cascade evolution.

\vspace{1mm}

\nin Let us discuss them briefly. ~~~The higher $\alpha_s$
perturbative terms can in particular essentially modify the
regular part of the BFKL-kernel \cite{fl}. But it probably will
not lead to a qualitatively new effects and can be taken into
account by  adjusting parameters of  ~\Pn trajectories and values
~\Pn of vertices. Already in the conformal approximation there are
the multigluon ladders \cite{bkp} which give additional regge
singularities with a vacuum quantum number. Probably every such an
object will also split into the complicated ensemble of discreet
levels (it will have the fine structure like the \Pn ), after
going from the conformal approximation by insertion of varying
$\alpha_s (u)$.
 The multigluon state can mix with \Pn state when one abandons the
conformal approximation, because the \Pn state ``consists'' of the
reggesed gluons, which by themselves are the multigluon  states.
Such a mixing may appear in higher in $\alpha_s (u)$ terms and the
corresponding vertices are of the same type as 3\Pm~ and higher
~\Pm~ vertices. Corresponding contributions are of approximately
the same form as the self-energy corrections to ~\Pm~ propagators
and vertices, and probably can be taken into account by  small
``phenomenological'' shifts of~$\Delta_n$.

The nonperturbative contributions to \Pm ~was partially discussed
in Introduction and it needs a separate detailed investigation.
Here we make only one note.
 At small $t$ there can be also essential the mixing of a $\Pn$
with states composed from light quarks in  specific configurations
corresponding to $2\pi$ states. This is connected to the existence
of the $\langle \bar{q}q \rangle$ condensate and manifests in the
large contribution of  $2\pi$ exchanges in multiperipheral
diagrams connected to pomeron. By this mechanism the intercepts of
the resulting states $j_n$ may be shifted on $\delta j_n (0) \sim
\pm \Lambda^2 < x_{\perp}^2>_n ~\sim \pm \exp{(-c n)}$. This shift
may be essential only for the first intercept $j_1(0)$,
representing the most soft part of pomeron. As a result it can
even be (and the data gives some support for this) that $j_1(0) <
j_2(0)$.

The other extension of BFKL is connected with fusion of gluons,
which becomes more and more essential at large $y$, and which
leads to the gluon saturation at different virtuality scales $u
\sim n$. The simplest model of gluon cascade describing all these
phenomena result from  the differential BFKL equation
(\ref{bfkl_dif}) supplemented by the nonlinear terms,
corresponding to 3\Pm ~(and all higher) diagrams
 \bel{nlin}
\frac{1}{\as(u)}\cdot \frac{\partial f(u,y)}{\partial
y}  ~=~ \delta~f + B \frac{\partial^2 f}{\partial u^2} - r(u)
f^2(u,y) +...,
 \ee
where $r(u)$ is proportional to the 3\Pm ~vertex at virtuality
$u$. The nonlinear corrections in (\ref{nlin}) correspond to
splitting and gluing of pomerons, and the perturbative in $r(u)$
solution for $f$ may be represented by the reggeon diagrams,
containing pomerons coming from the linear equation. A more
accurate than (\ref{nlin}) equations  \cite{jklw} contain in fact
the same physical information,~ and therefore we do not
need to enter here into details. \\

It seems therefore  plausible that  most higher $\alpha_s$
corrections, which in particular make amplitudes unitary in all
channels, may be taken into account by summing of contributions of
various reggeon diagrams with all \Pn states. The properties of
these diagrams can be summarized (as a simple generalization of
one pomeron case)  by the introduction of the effective Lagrangian
for the \Pn reggeon field theory
 \beal{rftl2}
 {\cal L} ~=~
  \sum_n  ~\Big(~ \Psi^+_n \frac{~\dlr }{2~\d y}\Psi_n  ~+~
  \Delta_n \Psi^+_n \Psi_n  ~~+~
  \alpha'_n ~\vec{\d_{\perp}} \Psi^+_n  ~\vec{\d_{\perp}}
  \Psi_n  ~+~~~~~~~~~~~~~~~~ \\
  ~+~(\Psi^+_n J_n + J^+_n \Psi_n)  ~\Big)
   ~~+~~
  i \sum_{m n k} ~r_{m n k} ~(\Psi^+_m \Psi^+_n \Psi_k +
  \Psi^+_k \Psi_m \Psi_n) ~~+ ~...~~~~~~~~~~~~\nn
\eea

\nin containing various \Pn interactions,  ~where $\Psi_n(y,
x_{\perp})$ ~is the ~\Pn -the reggeon field, $r_{m n k}$ ~-~
3~\Pm~~,... vertices between the corresponding \Pn states,~~ $J_n$
- are the external currents representing colliding
particles(nuclei), the simplest one are $J_n \simeq g_n^{(a)}
\delta (y-Y)~,~ J_n^+ ~=~ g_n^{(b )} \delta(y)$.~ From
(\ref{rftl2}) on can in usual way construct all reggeon diagrams
with \Pn.

The parton saturation in terms of reggeon diagrams corresponds to
transition to a state in which reggeon field operators
$\Psi_n(b,y)$  have the nonzero vacuum expectation value~$\langle
\Psi_n \rangle \sim \Delta_n/r_{nnn}$. ~The $\langle \Psi_n
\rangle$ condensation (saturation) takes place in the bubble that
grows with $y$ around external sources $J_n$. As a result the new
\Pn fields $\psi_n = \Psi_n - \langle \Psi_n \rangle$   enter into
the effective Lagrangian in such a $\langle \Psi_m \rangle$-bubble
with  different parameters $\Delta_n, \alpha'_n, r_{n~m~m}$, etc,
depending on the properties of this medium.

Such a behavior is typical for the supercritical reggeon theory
and leads to the Froissart asymptotics of cross-section. Here the
Froissart disk is a transverse region filled with condensed
$\langle \Psi(b,y) \rangle$. This picture can be generalized to
the multi -~\Pn reggeon field theory (see
Section 7).\\

To understand the range of applicability of the regge approach in
the case of supercritical pomeron one can estimate the average
rapidity intervals $\langle y_n \rangle$ on the ~$\Pm_n$ ~~lines
in general complicated reggeon diagrams, essential at asymptotic
energies, and find how $\langle y_n \rangle$ changes with growth
of the full rapidity $Y$. Its value is evidently connected with
amplitudes of pomerons splitting and joining defined by their
intercepts and values of $r_{n}$ vertices.

If we imagine such a reggeon as a particle propagating in a medium
composed from other similar pomerons, then a simple estimate of
time (rapidity) interval between interactions of this \Pn particle
is
 \bel{meany}
\tilde{y}_n \sim  \frac{1}{r_n^2 \rho_n}~~,
 \ee
where  $\rho_n \sim $ to a density of pomerons around the
considered pomeron, and entering relation $r_n^2$ is $\sim $ to
the probability the considered pomeron to interact with neighbor
pomerons in the unit of time $y$. In (\ref{meany})  we also
supposed that the main ~\Pn interactions are with pomerons of the
same size, and the 3\Pm ~vertex $r_n \simeq r_{nnn}$.

If the full rapidity interval $Y$ is so that we yet are far from
the \Pn saturation scale, the value of $\rho_n$ is small and it
gradually grows with $Y$. When the full energy riches the
saturation scale for \Pn~, then the mean \Pn density $\rho_n \sim
(\Delta_n/r_n)^2$, and it stops to grow with $Y$, so that for the
average  pomeron ``lifetimes'' in these conditions we have
$\tilde{y}_n \sim 1/\Delta_n^2$.

Because the average y-distance between the nearest steps on the
effective \Pn pomeron ladder is $\sim 1/\Delta_n$, the mean \Pn
pomeron line even in the saturated medium will have inside it
$\sim 1/\Delta_n \sim n/\alpha_s \gg 1$ ``ladder'' steps. And
therefore the \Pn quasi-particles can be consistently used also
inside the Froissart disk. \\

Now let us consider the multi-reggeon vertices with attached ~\Pn
~states.  At first sight the value of various multi-\Pm ~vertices
for ~\Pn may by estimated using vertices describing the joining of
many pomerons of the BFKL type at a given virtualities.

So, for example, if we consider the 3 \Pn vertex with all three $
n_1 \simeq n_2 \simeq n_3$ of the same order, then we have - $
\langle u_1 \rangle \simeq \langle u_2 \rangle \simeq \langle u_3
\rangle \sim n_i$.~ Then on dimensional grounds it may be expected
that $r_{3p}(u,u,u) \sim \Lambda_c^{-1} \exp (-u)$. This would be
correct if $r_{3p}$ represents the vertex in which one joins
pomeron states with definite $u_i$~, that is with definite
transverse sizes. But the \Pn states contain the superposition of
transverse distances with amplitudes $\psi_{n}$~, and therefore
$r_{3p} (n_1 n_2 n_3)$ is proportional to overlapping integrals
$$
\int du_1 du_2 du_3 ~\psi_{n1} (u_1) \psi_{n2} (u_2) \psi_{n3}
(u_3) ~\tilde{r}(u_1, u_2, u_3)~,
$$
where $\tilde{r}(u_1 u_2 u_3) \sim  \exp (- c u_i)$ corresponds to
the internal part of the $r_{3p}$ vertex. ~Because  $\psi_{n}(u)$
are not small at low $u$, it follows from (\ref{pvf}) that
$\psi_{n}(u \sim 1) \sim 1/\sqrt{n}$, ~so that the value of
$r_{3p}$ vertices can be estimated as
$$
r_{3p} (n_1 n_2 n_3) ~\sim~ \Lambda_c^{-1} ~(n_1 n_2 n_3)^{-1/2}
$$
All other mulipomeron vertices with definite $n_i$ have probably
the same structure - i.e~ are not too small.

But near the saturation region the value of vertices may
drastically change. So if the first $m$ states \Pn  are already
saturated the low $u$ part of all $\psi_{n}(u)$ are cut up to $u <
\exp (c m)$. And it will lead to the estimation
$$
r_{3p} (n_1 n_2 n_3 ; m) ~\sim \Lambda_c^{-1} \exp (- c m) ~.
$$

The regge description, especially with many \Pn , contains a large
number of  ``phenomenological'' parameters such as \Pn intercepts
$\Delta_n$ and slopes $\alpha'_n$, entering \Pn trajectories
$j_n(t)$, the \Pn interaction vertices $r_{mnk}$, and the
inclusive vertices $\gamma_n$.~~  In principle all these
quantities  ``may'' be calculated directly from QCD as a function
of $\alpha_s$ and some nonperturbative QCD parameters. But now it
is probably too early to hope that it can be done in a
quantitative way.

\subsection*{~5.~~ \Pn ~states and the parton picture }
\vspace{2mm} \nfn{5}

The parton picture corresponding to BFKL pomeron is relatively
simple. The mean parton configurations for a fast colorless hadron
are produced by the branched gluon cascade in $y$ with average
number of steps $\sim \alpha_s y$. This leads to an exponential
growth with $y$ of the number of low energy partons(gluons).

In the case of not running $\alpha_s$ the parton ``motion'' in its
virtuality $u = \log k^2_{\bot}/\Lambda^2$  is
conformal-invariant, and at growth of $y$ the free random motion
of partons virtuality $u_i(y)$ takes place. As a result the
lowermost partons have the virtualities $u_i \sim \sqrt{\alpha_s
y}$ on average.
\\

\begin{figure}[h]
\begin{center}
\includegraphics[scale=0.7,keepaspectratio=true]{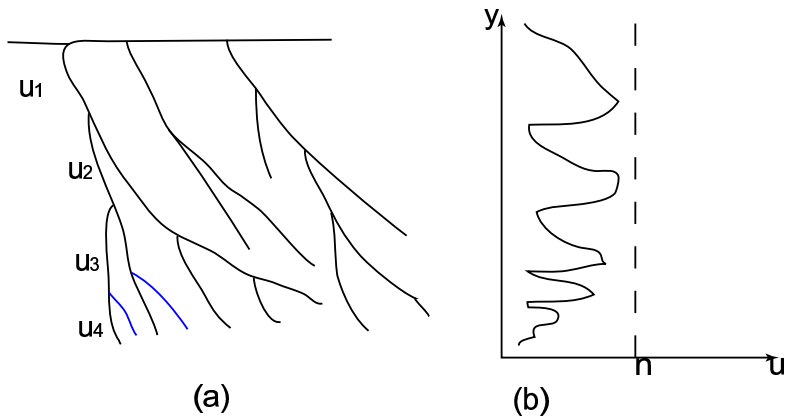}
\caption{\small  (a)~~Parton state - Gluon cascade,
  with dedicated parton branch.
 (b)~ The u-motion of partons on the selected parton branch. }
 \label{Fig1}
\end{center}
\end{figure}

In the case of running $\alpha_s (u)$ the diffusion in $u$ takes
place with the additional drift in the direction of small $u$. The
corresponding distribution in virtuality of bottom-partons may be
represented as the approximate solution of Eq.(\ref{bfkl_dif}) in
such form:
 \bel{u-distr}
 f(u,y) ~\sim~ \frac{f_0}{y}~\exp{\Big(~
 \delta~y~\alpha_s(u)
 - \frac{(u-u_0)^2}{4~B~y~\alpha_s(u)} ~\Big)}~~,~~~~~
 \alpha_s(u) = \frac{1}{a+bu}~~~~
 \ee
From (\ref{u-distr}) it evidently follows that at large $y$ the
mean virtuality is
$$
<u> ~=~ \nin  \int u f(u,y)du ~\Big/ ~\int
        f(u,y)du \sim const(y)   \sim 1~,
$$
and does not grow with $y$. But the mean $\langle  k^2_{\bot}
\rangle$ will grow as $ \exp{ ( 2\sqrt{y B/b})}$, and this
reflects the u-motion of partons in the ~\Pn states .

A \Pn state does not directly correspond to a definite parton
configuration of fast hadron, but only to a superposition of
specific choices of parton branches which interact with a target
 \footnote{Here the situation is similar to the case for other
regge poles, for example, non-vacuum, in that case the interaction
with the target ``selects'' the branch of a parton cascade through
which the corresponding quantum number is transported from the
valence quark region.}.
 The \Pn approximately corresponds to parton branches interacting
with the target with the mean bottom virtuality $u \sim n$. For a
large $y$ the value of virtuality $u_i(y)$ along such a parton
branch  ``oscillates'' between $u \sim 1$ and $u \sim n$. The
period in $y$ of such $u$-oscillations is $\tau_n \simeq 3n^2$.
These values directly correspond to the splinting between adjacent
\Pn levels $\tau_n^{-1} \simeq |\Delta_n - \Delta_{n+1}|$.

From here it follows some evident limitation for using \Pn states
in phenomenology, because $\tau_n$ grow rather fast, and the
currently experimentally reachable rapidities are limited $y < 20
\div 25$. So only the first few \Pn states can be distinguished,
and all higher ones act as a single hard regge singularity at $j
\simeq 1$.

The other limitation is connected with the fusion of gluons, which
becomes more and more essential at large $y$, and the transition
to the gluon saturation at the different virtuality scales $u \sim
n$. The simplest model of gluon cascade describing all these
phenomena results from  the differential BFKL equation
supplemented just as in (\ref{nlin}) by the nonlinear terms.

Neglecting in  (\ref{nlin}) the diffusion term $\partial^2
f/\partial u^2$, we become the simple solution
 \beal{nlin-s}
 f(u,y) ~=~ \frac{ \exp \big(\delta y \alpha_s(u)\big)
   }{f_0(u)^{-1} ~+~ \delta^{-1}~r(u)
   ~\Big(\exp \big(\delta y \alpha_s(u)\big) -1\Big)}
~~\Longrightarrow ~~~ \frac{\delta}{r(u)}~, \\
f_0(u) ~=~ f(u,y=0)~~~~~~~~~~~~~~~~~~~~~~~~~~~~~~~~~~~~~~~~~~\nn
 \eea
at $y \Longrightarrow\infty$, so that for every virtuality the
gluon density stops to grow after rapidity reaches some critical
value depending on $u$. This parton saturation for the virtuality
$u \sim n$ takes place at
$$
 \tilde{y} ~\geq~ \frac{1}{\delta \alpha_s(u)}
\log \frac{\delta}{f_0(u) r(u)} ~~\sim~
         \frac{b}{\delta}~u^2 +  u c(u)
~~~,~~~u \simeq n
$$
The last expression is the estimation for large $u$, where we
supposed that $\log r(u)^{-1} \sim u$. So with growth of $y$ the
consecutive saturation of parton densities at higher and higher
scales $\tilde{u}(y) \sim \sqrt{y\delta/b}$ takes place, and it
may be interpreted as a consecutive saturation of densities of the
\Pn pomerons with $n \sim \tilde{u}(y)$.

\vspace{2mm}

\subsection*{\bf 6. Regge phenomenology  with~\Pn} \vspace{2pt}
%
The contribution of higher \Pn states may be essential for a
description of various reactions in the regge approach, especially
in the cases when some of the transverse momentum transfers become
large. But, as it was mentioned above, the separation of
contributions of \Pn at ``real'' and not at asymptotic $y
\rightarrow \infty$ energies cannot be done. So, possibly, at all
experimentally accessible energies the reasonable  approximation
may contain a short sequence of distinguishable \Pn states
$n=1,2,...m$, $m \sim 2 \div 3$ and the rest \Pn ~$n>m$ will look
as one indistinguishable hard pomeron singularity $\tilde{\Pm}$ at
$j=1$.

It seems that one can base a reasonable phenomenology for all main
high energy processes using such a truncated  system of \Pn
states. It can look much simpler than approaches based on a direct
use of QCD variables. From one side, it hides the really essential
degrees of freedom, but from the other side, the unitarity
conditions in all channels are explicitly under control.

In the next Section we consider briefly the asymptotic Froissart
behavior to illustrate the process of parton saturation at various
virtuality scales, and how it arises from \Pn states.

\subsection*{\bf 7.~~Saturation and Froissart behavior}
\nfn{7} \vspace{1pt}

The Froissart type asymptotics always is expected \cite{FroiDisk}
for supercritical reggeon theories - in QCD we have directly this
case.~~ The usual way to  such a behavior comes from an eikonal
summation of pomeron exchanges, and it leads to the inelastic
cross-sections $\sigma(y)=\simeq \pi R^2(y)$. In this case the
radius of the soft Froissart disk in the impact parameter space
$b_{\bot}$ grows with $y$ like $R(y) = v y$, where the transverse
``velocity'' of growth $v = 2\sqrt{\alpha' \Delta}$. This
mechanism is rather primitive - pomerons do not interact one with
another - and as a result the effective parton (and pomeron)
density exponentially grows with $y$ inside the ``disk'', and we
have not a saturation but only a screening of partons. But
occasionally this simple model gives many correct predictions. It
leads to a black and not a gray Froissart disk, and  the
corresponding parton model is boost-invariant. Probably, the main
defect of this model is that it is soft, but this disadvantage may
be cured if we generalize the model to the multi~ \Pn case and
take into account the pomeron interactions which lead to the
parton saturation.

This looks quite easy.  Taking for the \Pn contributions in the
usual factorized form
 \bel{npol}
 \chi_n (b_{\bot},y) ~=~
   i~\frac{g_n \cdot g_n}{\alpha^{\prime}_n y_1}
~\exp \Big(~ \Delta_n y_1-b^2_{\bot} / 4\alpha^{\prime}_n y_1
~\Big) ~,
 ~~y_1 = y - i \pi \Delta_n/2~,
 \ee
we have for the full $S(b_{\bot},y)$-matrix and amplitude the
eikonal like
 \footnote{Note that there are no reasonable arguments that the
full contribution from the exchange of noninteracting \Pm ~should
take the eikonal form $S = \exp{i \chi}$. The correct form of
$S[\chi_n]$ may be completely different because the contribution
from  diffractive jets may be very big; for example $S$ can behave
like $(1 - i \chi)^{-1}$. But this will not change the general
structure of the Froissaron - the full picture of embedded disks
remains the same.}
 expressions
 \beal{sm}
 S(b_{\bot},y) ~=~  S[\{\chi_n \}] ~=~
   \exp \Big( ~i\sum_n \chi_n (b_{\bot},y)\Big)~\simeq~
         \theta (b^2 - R^2_1(y))~,\\
 A(b_{\bot},y) ~=~ i(1 - S(b_{\bot},y))  ~\simeq~
    i\theta (R^2_1(y)-b_{\bot}^2)~,~~~~~~~~~~~~~~~~~~~~~~~
\nn
 \eea
where $ R_1(y) = y \sqrt{\alpha'_1 \Delta_1}$ is the soft disk
$\F_1$ radius.
 This soft disk contains also a chain of more hard disks
$\F_n$  with smaller radii ~$R^2_n(y)~=~ \alpha'_n \Delta_n y^2$
$\equiv v_n^2 ~y^2$ and large average parton(pomeron) virtuality
$u_n \sim n$. The hard disks  $\F_n$ are only slowly reflected in
total cross sections - they simply make the internal part of the
Froissart disk more dark. But if we consider the inclusive
cross-section for particles with a high mass or other events with
high virtuality the created particles come mostly from the
corresponding hard disks $\F_n$. This in particular leads to the
growth of mean transverse momenta of secondary particles with
energy.

Let us examine how this simple picture changes when we take into
account the \Pn pomeron interactions. Consider the soft disk
$\F_1$ composed from $\Pm_1$. ~Near the border of $\F_1$ at ~$b
\simeq v_1 y$ ~the $\Pm_1$ ~density is $\sim 1$, and here the
reggeon interaction is not especially essential. ~Just the rate of
parton splitting in this region defines velocity $v_1$ of
$\F_1$-disk growth with $y$. Therefore in the first approximation
the effect of pomeron fusion are essential only in the inner parts
of $\F_1$-disk where we may also neglect the transverse (in
$b_{\bot}$) motion of $\Pm_1$-reggeons and use for its density the
expression (\ref{nlin-s}). This gives the value of the saturated
~$\Pm_1$ density inside $\F_1$ ~of order ~$f_1 \simeq
\Delta_1/r_1$. ~It must be corrected only near the border of
$\F_1$ where density changes from $f_1$ to values of the order
$\sim 1$. The width of this strip is $\sim
\sqrt{\alpha'_1/\Delta_1} \log (\Delta_1/r_1)$.

Now consider the higher $\F_n$-disk and ask by what a way the
density of \Pn at large distance $b_{\bot}$ is generated. Various
mechanisms may operate. The one is connected with the direct
growth of $\F_n$ disk from $b_{\bot} =0$, but it is very slow and
leads to the small radius of saturated disk~ $R_n \sim y /n^3$.
The most effective mechanism \cite{KanchFroi} that transports the
high virtuality partons to larger $b_{\bot}$ is connected with the
local growth of $\F_n$ density from the soft saturated $\F_1$ disk
whose radius has already reached this value of $b_{\bot}$. To
describe it we may use the expression (\ref{nlin-s}) for $f(u,y)$
where we change $y \rightarrow y - b_{\bot}/v_1$ and use for
$f(u,0) \sim \exp(-u)$. From this we conclude that the saturation
at scale $u$ and distance $b_{\bot}$ is reached at $y =
b_{\bot}/v_1 + (2b/\delta) u^2$. This corresponds to the saturated
$\F_n$ disk radius
$$
 R_n(y) ~\sim~ v_1 y -  \langle u^2 \rangle_n ~(2 b v_1/\delta) ~=~
             R_1(y) - n^2 \lambda~,~~~~
\lambda = 2v_1 b c_3/\delta
$$
It follows from here that the maximal saturated virtuality in the
center of \F disk is  $u  \sim \sqrt{y}$.  One can also define
approximately the width $\delta R_n$ of the $\F_n$ border from the
condition that the reggeon density changes from the saturated
value $f_n \sim \Delta_n / r_n$ up to $\sim 1$. From this we
become $\delta R_n \sim n~v_1 \Delta^{-1}$.

Note that the value of the transparency of full \F disk should not
depend essentially from the pomeron interactions. This is because
the main contribution to \F -transparency, which is $\sim
|S(b_{\bot},y)|^2$~,~ comes from the components of the fast hadron
wave function  without a parton cloud and correspondingly without
\F disk. An estimation of this probability gives the
boost-invariant answer $|S(b_{\bot},y)|^2 \sim \exp
(-c(b_{\bot})~y)$. ~~This quantity enters various cross-sections,
and it would be interesting to check its properties experimentally
at the most possible energies, because this may reflect essential
elements of parton dynamics.

\vspace{4mm}

\section*{\bf Conclusion}
\nfn{8}
The aim of this article  was to discuss  the main physical
properties of \Pn states and to understand if the use of \Pn in
regge phenomenology may be natural and useful. ~Our answer is more
positive than negative ~~It seems that the approach based on the
multi-\Pn sequence may adequately represent most essential aspects
of high energy interactions - from QCD perturbative to
nonperturbative ones, in particular a growth of a mean virtuality
with energy and parton density saturation. Also the purely
asymptotical picture of the Froissart limit as represented by a
system of nested disks filled by the \Pn condensate looks quite
reasonable.

It is essential that by using the regge approach with multiple \Pn
, instead of working directly QCD degrees of freedom,  we may
expect that the unitarity restrictions on  amplitudes from
different crossing channels are almost automatically taken into
account. This is a big advantage from the phenomenological point
of view. But even at the fundamental theoretical level it may be
more simple to calculate once the values of regge vertices for
~\Pn~ from QCD, than do all this from the beginning for every
amplitude.

\vspace{12mm}
\nin {\bf ACKNOWLEDGMENTS} \\
\nin I would like to thank K.G.~Boreskov for conversations and
comments.~~

\nin The financial support of CRDF through the grants
RUP2-2961-MO-09 is gratefully   acknowledged. 
\end{document}